\documentclass[prb,showpacs,floatfix,amsmath,amssymb,superscriptaddress]{revtex4-1}
\usepackage{amsfonts}
\usepackage{stmaryrd}
\usepackage{bbm}
\usepackage{mathrsfs}
\usepackage{tipa}
\usepackage{amssymb}
\usepackage{txfonts}
\usepackage{graphicx}
\usepackage{dcolumn}
\usepackage{epstopdf}
\usepackage[colorlinks,linkcolor=blue,urlcolor=blue,citecolor=blue]{hyperref}
\usepackage{multirow}
\usepackage{subfigure}
\usepackage{url}
\usepackage{upgreek}

\begin{document}

\newcommand*{\cm}{cm$^{-1}$\,}
\newcommand*{\Tc}{T$_c$\,}


\title{Light induced sub-picosecond topological phase transition in MoTe$_2$}

\author{M. Y. Zhang}
\affiliation{International Center for Quantum Materials, School of Physics, Peking University, Beijing 100871, China}

\author{Z. X. Wang}
\affiliation{International Center for Quantum Materials, School of Physics, Peking University, Beijing 100871, China}

\author{Y. N. Li}
\affiliation{International Center for Quantum Materials, School of Physics, Peking University, Beijing 100871, China}

\author{L. Y. Shi}
\affiliation{International Center for Quantum Materials, School of Physics, Peking University, Beijing 100871, China}

\author{D. Wu}
\affiliation{International Center for Quantum Materials, School of Physics, Peking University, Beijing 100871, China}

\author{T. Lin}
\affiliation{International Center for Quantum Materials, School of Physics, Peking University, Beijing 100871, China}

\author{S. J. Zhang}
\affiliation{International Center for Quantum Materials, School of Physics, Peking University, Beijing 100871, China}

\author{Y. Q. Liu}
\affiliation{International Center for Quantum Materials, School of Physics, Peking University, Beijing 100871, China}

\author{C. N. Wang}
\affiliation{International Center for Quantum Materials, School of Physics, Peking University, Beijing 100871, China}

\author{Q. M. Liu}
\affiliation{International Center for Quantum Materials, School of Physics, Peking University, Beijing 100871, China}

\author{J. Wang}
\affiliation{International Center for Quantum Materials, School of Physics, Peking University, Beijing 100871, China}
\affiliation{Collaborative Innovation Center of Quantum Matter, Beijing, China}

\author{T. Dong}
\email{taodong@pku.edu.cn}
\affiliation{International Center for Quantum Materials, School of Physics, Peking University, Beijing 100871, China}

\author{N. L. Wang}
\email{nlwang@pku.edu.cn}
\affiliation{International Center for Quantum Materials, School of Physics, Peking University, Beijing 100871, China}
\affiliation{Collaborative Innovation Center of Quantum Matter, Beijing, China}

\begin{abstract}
Recent development of ultrashort laser pulses allows for optical control of structural and electronic properties of complex quantum materials. The layered transition metal dichalcogenide MoTe$_2$, which can crystalize into several different structures with distinct topological and electronic properties, provides possibilities to control or switch between different phases. In this study we report a photo-induced sub-picosecond structural transition between the type-\uppercase\expandafter{\romannumeral2} Weyl semimetal phase and normal semimetal phase in bulk crystalline MoTe$_2$ by using ultrafast pump-probe and time-resolved second harmonic generation spectroscopy. The phase transition is most clearly characterized by the dramatic change of the shear oscillation mode and the intensity loss of second harmonic generation. This work opens up new possibilities for ultrafast manipulation of the topological properties of solids, enabling potentially practical applications for topological switch device with ultrafast excitations.

\end{abstract}

\maketitle

Layered transition metal dichalcogenide (TMD) MX$_2$ (where M is a transition metal (Ta, Nb, Mo, and W) and X is a chalcogen (S, Se, and Te)) represents one of the most interesting material groups that shows rich physical phenomena such as charge density wave, superconductivity, quantum spin Hall effect, topological nontrivial state, unsaturated extremely large magnetoresistivity (XMR), etc. Among them, MoTe$_2$ has recently attracted particular attention because it hosts type II topological Weyl fermions in a noncentrosymmetric octahedral T$_d$  structural phase \cite{Soluyanov2015,Sun,Chang2016}, which allows for fundamental studies of intriguing topological phyiscs. Actually, MoTe$_2$ can crystallize into several different structures with different properties, for example trigonal prismatic coordinated hexagonal 2H phase (space group P6$_3$/mmc), distorted octahedral coordinated monoclinic 1T$^{'}$(space group P2$_1$/m) and orthorhombic T$_d$ (space group Pmn2$_1$) phases \cite{Dawson,Clarke,Nature2017}. The 2H phase has a central symmetric structure and is semiconducting, but 1T$^{'}$ and T$_d$ phases show semimetallic nature. The 1T$^{'}$-MoTe$_2$ phase is centrosymmetric and can be transformed into T$_d$-MoTe$_2$ phase with broken inversion symmetry by cooling the compound below $\sim$250 K \cite{Dawson,Clarke}. T$_d$-MoTe$_2$ is also found to be superconducting with T$_c$=0.1 K, thus serving as a promising candidate of the topological superconductor \cite{SC,Takahashi,doi:10.1063/1.4947433}. Since the energy difference between different phases is small in MoTe$_2$ \cite{Kim}, the system provides a good possibility to manipulate or engineer controlled phase transitions.

The lattice structures and symmetries of materials are usually probed and determined by elastic X-ray, neutron and electron diffractions or inelastic Raman spectroscopy. With the development of advanced femtosecond laser techniques, ultrafast coherent phonon spectroscopy based on pump probe technique and second harmonic generation (SHG) spectroscopy provide alternative routes to resolve lattice symmetry\cite{Huber,Wall,Beaud1,Beaud2,PRLBi2Se3,BykovBi2Se3,GaAs2003}. Such optical probes are particularly effective for detecting structural change associated with inversion symmetry breaking. When the inversion symmetry is broken, parity is not a good quantum number, giving rise to a mixture of even and odd states. The sudden lattice symmetry change would lead to a variation in A$_1$ Raman active phonon modes, which can be tracked by the ultrafast coherent phonon spectroscopy through the displacive excitation of coherent phonons or stimulated Raman scattering process\cite{DECP,ISIS}. SHG spectroscopy is extremely sensitive to inversion symmetry because the leading order contribution to SHG signal is nonzero only when the inversion symmetry is broken, e.g. at the interfaces or in bulk non-centrosymmetric crystalline. More importantly, the employment of such optical probes allows for time resolved measurement after ultrafast photo excitations.

Up to now, the experimental realization of switch between different structural phase transitions in MoTe$_2$ was mainly achieved via application of strain, doping or static electric field on few layers samples \cite{Wang,Sakaie1601378,Zhang}. Ultrashort laser pulses provide a new route to manipulate the structural and electrical properties of a quantum material. When the electric field of the pumping pulse is comparable to the potential gradients of atoms, the intense pulse can change the landscape of free energy and induce a phase transition. There are many good examples of the ultrafast photoinduced transitions such as the insulator-to-metal transition in VO$_2$ \cite{PhysRevLett.87.237401,Wall}, the lattice distortions in manganites\cite{Rini2007b} and the charge density wave to metal transition in  K$_{0.3}$MnO$_3$\cite{Huber,DemsarKMnO}. A recent study indicated that a strong laser radiation can induce a non-reversal structural phase transition between 2H and 1T phases in MoTe$_2$ \cite{radiation}. Here we report an ultrafast photoinduced sub-picosecond topological structural transition between the type-\uppercase\expandafter{\romannumeral2} Weyl semimetal T$_d$ phase and normal semimetal 1T$^{'}$ phase in bulk crystalline MoTe$_2$. The onset and degree of the phase transition are determined by the transient photo-induced coherent phonon spectroscopy and time-resolved SHG separately.

\textbf{Results}

\begin{figure*}
  \centering
  \includegraphics[width=15cm]{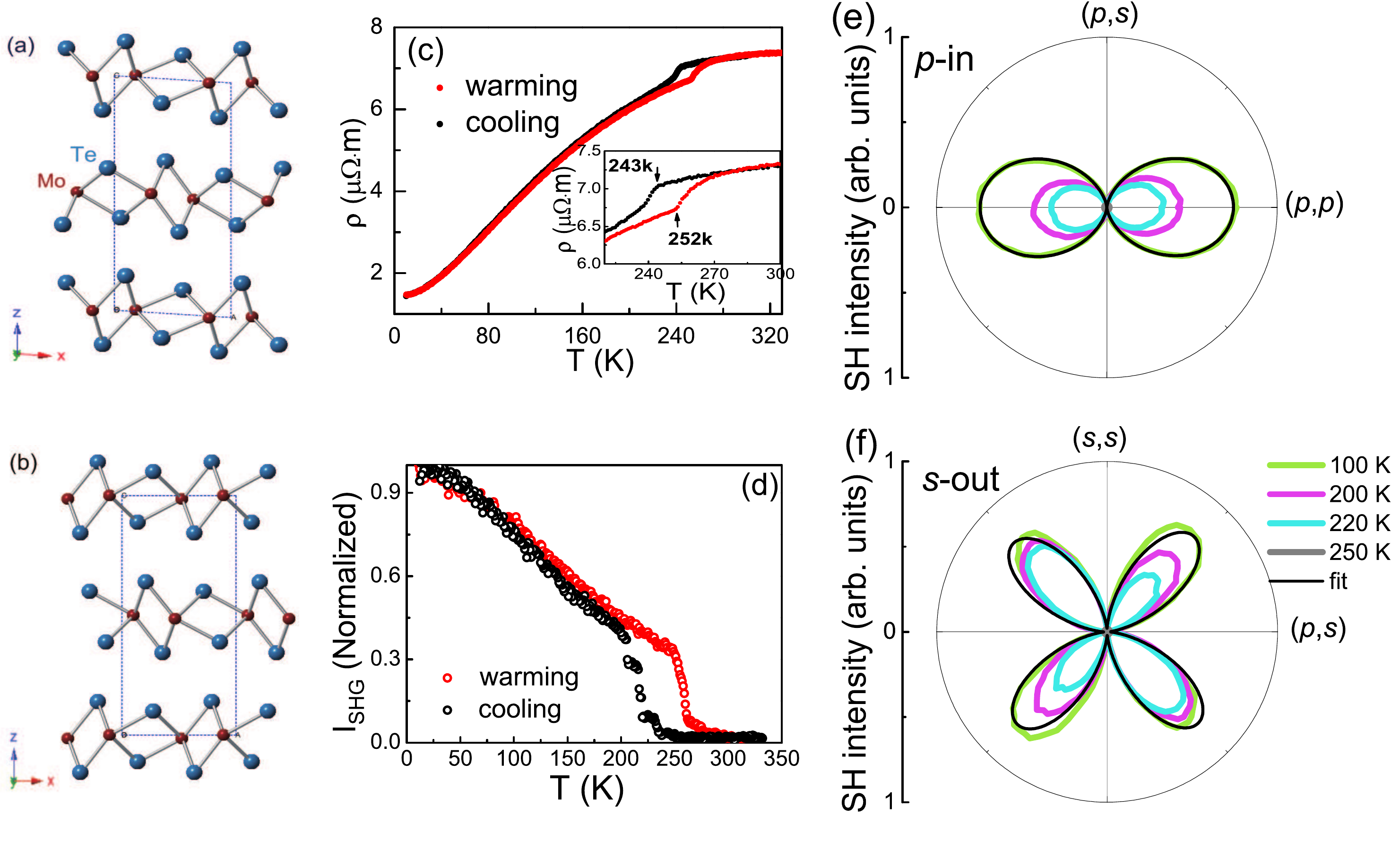}
  \caption{ Sample structure and characterization. Crystal structures of MoTe$_2$ in (a) 1T$^{'}$ and (b) T$_d$ phases. Both (c) Resistivity measurement and (d) the optical SHG intensity measured with an incident photon energy of 800 nm show a temperature-induced first order phase transition. SHG characteristics (e) for the incidence p-polarized while the SH analyzer is rotated and (f) for the s-polarized SH photons while the incidence polarizer is rotated.}\label{Fig:1}
\end{figure*}

\textbf{Sample characterization}. We perform measurement on the 1T$^{'}$-MoTe$_2$ single crystals which have a first order structural transition (T$_V$) near 250 K. The high temperature centrosymmetric monoclinic 1T$^{'}$ structure and low temperature noncentrosymmetric T$_d$  structure are shown in the Fig.\ref{Fig:1} (a) and (b), respectively. The structural difference between the two phases is small, only by a 4$^\circ $ tilting angle from the stacking direction (c-axis). The type-II Weyl semimetal characteristics in T$_d$ phase were identified by angle resolved photoemission spectroscopy and scanning tunnelling spectroscopy studies\cite{NM2016,NP2016}. It should  be pointed out that the single layer of the two phases is the same and they differ only in the layer stacking. Figure \ref{Fig:1} (c) shows the temperature-dependent resistivity of the crystal. Presence of significant hysteresis in transition temperatures between cooling down (about 240 K) and warming up (about 250 K) (inset of Fig.\ref{Fig:1} (c)) yields evidence for the first order phase transition at T$_V$.  Figure \ref{Fig:1} (d) shows the optical SHG intensity at a particular measurement configuration of p-polarized incidence and p-polarized second harmonic (SH) photons. The signal is absent at high temperature but increases suddenly below the phase transition, yielding evidence for the inversion symmetry breaking in the T$_d$ phase. The signal also shows a clear hysteresis across T$_V$ with cooling down and warming up the sample, further supporting the first order phase transition. Figure \ref{Fig:1} (e), and (f) show the SH polarization patterns in MoTe$_2$ for two different measurement configurations, with a fixed p-polarized incidence-rotatable SH analyzer and a fixed s-polarized SH photons-rotatable incidence polarizer, respectively. The optical setup is described in detail in the methods. The results are similar to the earlier report based on the same experimental configuration \cite{Sakaie1601378}. The polarization patterns can be well reproduced by the symmetry analysis based on the space group Pmn2$_1$ (see details at the symmetry analysis in Methods).

\begin{figure*}[htbp]
  \centering
  \includegraphics[width=12cm]{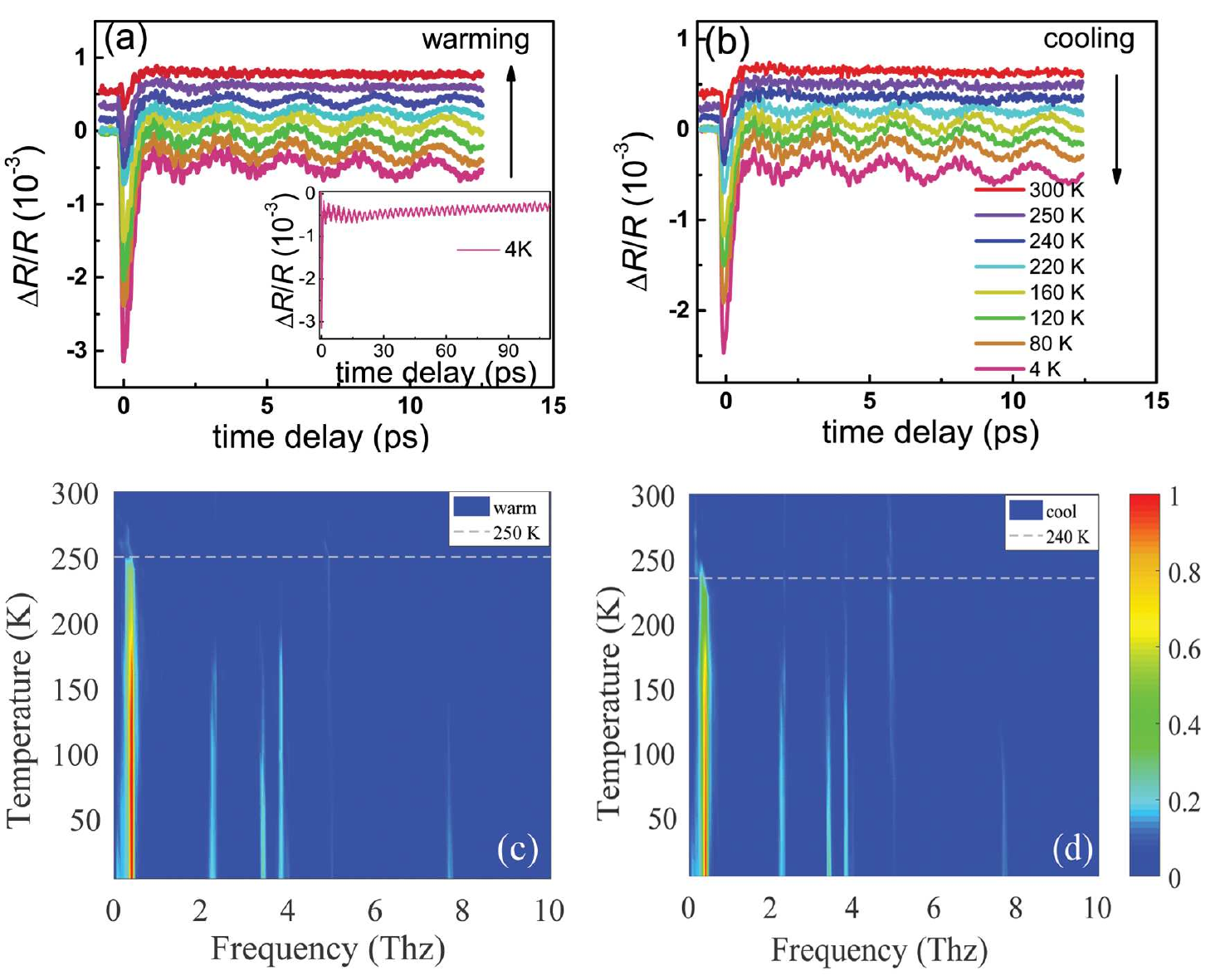}
  \caption{The temperature-dependent coherent vibrational dynamics. The transient pump-induced reflectivity at 800 nm (a) from 4 K to 300 K (warming process) and (b) from 300 K to 4 K (the cooling process). Above 240 K, all the curves are vertically shifted for clarity. Inset of (a):  Pump-induced reflectivity change at a longer time delay scale at 4 K. We can see that the coherent phonons last for very long time delay. Color plots of the Fourier transformation (FT) of the transient reflectivity after subtracting the background: (c) the warming process and (d) the cooling process at selected temperatures.}\label{Fig:2}
\end{figure*}

\textbf{Ultrfast coherent phonon spectroscopy.} Figure \ref{Fig:2} (a) and (b) show the transient reflectivity of MoTe$_2$ measured from 4 K to 300 K (the warming process) and from 300 K to 4 K (the cooling process) separately using a Ti:sapphire amplified laser at low influence (see method). The decay dynamics contains two processes: a fast component ($\tau_1$) about 0.3 ps related to the intraband transition and a slow component ($\tau_2$) about hundreds of picoseconds, which are similar to other TMD materials \cite{PhysRevBWTe2,MoS2}. The coherent vibrational dynamics is obtained after subtracting the two decay processes. The coherent vibrational dynamics contains the oscillation signals originating from the low energy A$_1$ optical phonons\cite{DECP}. It can be seen clearly in the time-domain spectrum that a long-period optical coherent phonon suddenly disappears above the structural phase transition temperature.

Figure \ref{Fig:2} (c) and (d) respectively show a Fourier transform of these oscillations of Fig.\ref{Fig:2} (a) and (b) after subtracting the two-exponential decay components of the form $\Delta R/R$ = $A_0$+$A_1$ exp$(-t/\tau_1)$+$A_2$ exp$(-t/\tau_2)$, where $A_i$ and $\tau_i$ (i = 1, 2) are the fit parameters that represent the amplitude and recovery rate of the dynamics, respectively. Six distinct peaks are observed in the Fourier spectrum at 0.42, 2.3, 3.4, 3.89, 5.0 and 7.77 THz ($\sim$14, 77, 113, 130, 167 and 259 cm$^{-1}$) at the low temperature. We found all those phonons have been detected in the Raman spectrum as A$_1$ modes \cite{Raman,NLRaman,Beams2016}. The 0.42 THz phonon, which appears around 240 K during the cooling process (Fig.\ref{Fig:2} (d)) and disappears around 250 K during the warming process, has much stronger intensity (Fig.\ref{Fig:2} (c)). It shows the same hysteresis as observed in the temperature dependent resistivity and SHG signals. The same hysteresis has also been reported in the Raman measurement\cite{Raman,NLRaman}. Based on symmetry analysis and calculation results reported before, the 0.42 THz phonon is a shear mode between layers and is Raman active in T$_d$ phase and inactive at 1T$^{'}$ phase (since the zone-center lattice vibrations have odd parity in the inversion symmetric structure at 1T$^{'}$ phase) \cite{NLRaman}. We can take this shear mode at 0.42 THz as an indication for the change of the structural transition in this compound. In principle, the other $A_1$ phonon modes should also change at the phase transition. However, those phonon modes have much weaker signal levels. Furthermore, since the structure change is very subtle and related modes at similar frequencies are also present at 1T$^{'}$ phase based on calculations \cite{Raman,NLRaman,Beams2016}, their change could not be well resolved. So, we shall not discuss them below.

\begin{figure}[htbp]
  \centering
  \includegraphics[width=12cm]{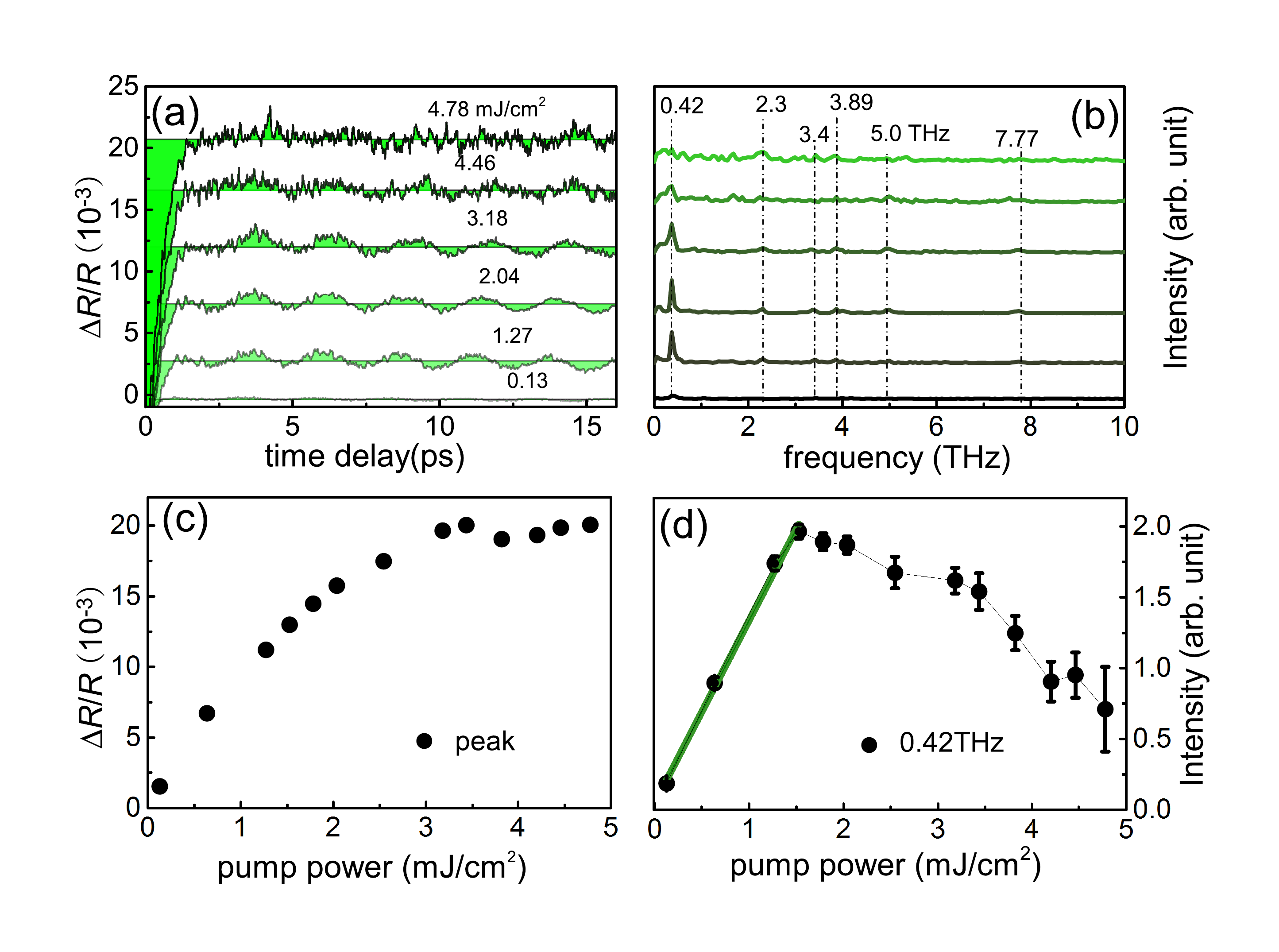}
  \caption{The Fluence-dependent coherent vibrational dynamics. (a) The transient pump-induced reflectivity changes in MoTe$_2$ at 4 K with different excitation fluences. (b) The Fourier transformation (FT) of the transient reflectivity after subtracting the background in (a). (c) The fluence dependence of the maximum change of the reflectivity. (d) The fluence dependence of the amplitude of the phonon at 0.42 THz.}\label{Fig:3}
\end{figure}

Having identified the shear mode as an indication of structural transition\cite{Raman,NLRaman,Zhao}, we now explore the fluence dependence of pump-probe signal. Figure \ref{Fig:3} (a) presents the pump-induced reflectivity change of MoTe$_2$ for different selected pump fluences at the low temperature T=4 K. The curves are vertically shifted for clarity. The Fourier transformations of the oscillations after subtracting the background are shown in Fig. \ref{Fig:3} (b). Under very low fluence, 0.13$<$F$<$1.5 mJ/cm$^2$, the signal is so small that only the strongest 0.42 THz phonon can be detected. With increasing the fluence, all other five phonons show up but the 0.42 THz phonon tends to disappear near 5 mJ/cm$^2$. Figure \ref{Fig:3} (c) displays the fluence dependence of the maximum change of the reflectivity  (peak intensity in Fig. \ref{Fig:3} (a)). It is clearly seen that the maximum change saturates at high fluence. Figure \ref{Fig:3} (d) shows the fluence dependence of the amplitude of the shear mode phonon at 0.42 THz. It increases linearly at small fluence, but decreases after reaching a maximum near 1.5 mJ/cm$^2$. According to the theory of  displacive excitation of coherent phonons\cite{DECP}, the intensity of the coherent phonon should increase with increasing fluence before reaching the damage threshold of a material \cite{bismuth1996}. There could be two possible reasons for the intensity reduction of 0.42 THz phonon mode in the high fluence. One is a de-phasing mechanism arising from oscillation damping and another is due to the symmetry change of lattice structure. Since the coherent phonon manifests as an oscillation in the recorded time trace of transient reflectivity, we can roughly estimate de-phasing time from the damping of oscillation. As shown in Fig. \ref{Fig:3} (a), the oscillation still lasts for long time at the high pump fluence, suggesting an almost unchanged lifetime of the coherent phonon, which essentially excludes the de-phasing mechanism. Then we need to consider the pump-induced structural symmetry change. The compound has a first-order transition at about 250 K and the 0.42 THz phonon mode disappears in the high temperature phase. A possible scenario is that the intensive pump can directly drive a structural change in a way towards its high temperature phase. In fact, photoinduced structural phase transitions have been found in different compounds \cite{Rini2007b,VanceR2014,Gedik2007}. Another possibility is the thermal effect in lattice temperature induced by the intense pump \cite{PhysRevLett.120.047601,WTe2}. In the latter case, the expected temperature rise $\Delta T$ at temperature ($T_0$) can be estimated by the equation $S\sigma \rho /M\int_{T_{0}}^{T_{0}+\Delta T}C_{p}(T)dT=(1-R)FS$, where $S$, $\rho$, $M$, $R$, $F$ and $\sigma$ are the excitation area, the mass density (7.78 $g/cm^3$), the molar mass (351.1 $g/mol$), the reflectivity of MoTe$_2$ (about 0.45 at 800 nm from our own measurement), the laser fluence and the optical penetration depth (about 100 nm if assuming a similar value as WTe$_2$ \cite{WT2-spectrum}), respectively. $C_{p}$ is the temperature-dependent thermal capacity. We estimate the values of $\Delta T$ to be $\sim$ 120 K and $\sim $ 300 K with excitation fluence about 1 $mJ/cm^2$ and 5 $mJ/cm^2$ respectively at $T_0$=5 K, which is also in agreement with the lattice thermal effect. Future efforts are needed to resolve the two scenarios.

\begin{figure}
  \includegraphics[width=15 cm]{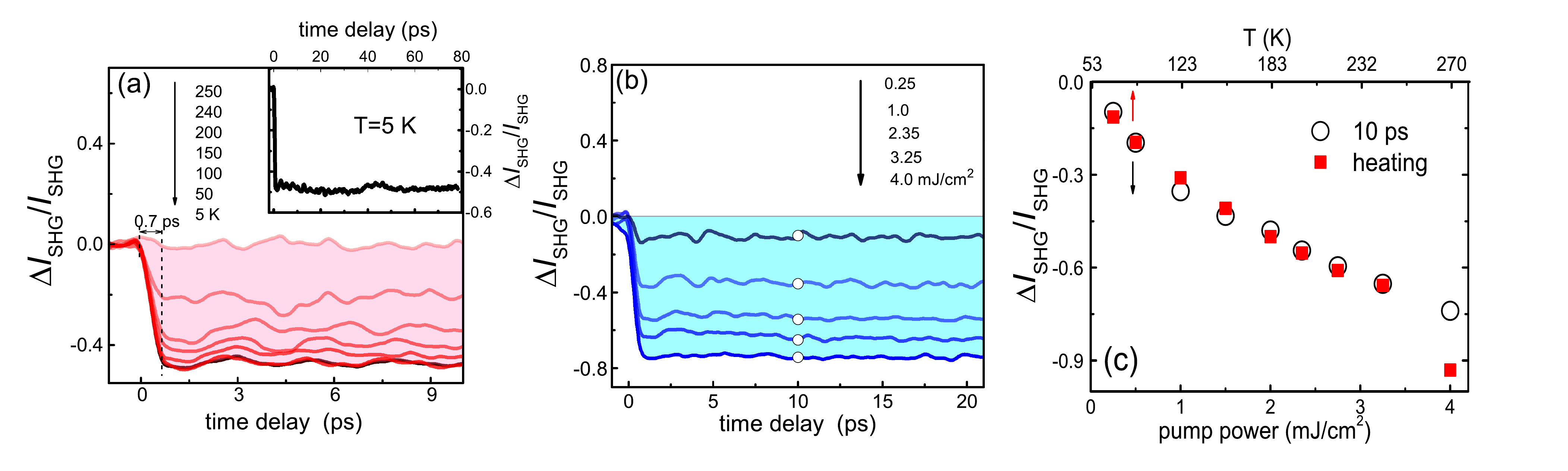}
  \caption{(a) The relative change of SHG at fixed pump fluence about 2 $mJ/cm^2$ measured at different temperatures. The inset is the signal at longer time delay. (b) The relative change of SHG as a function of time delay with selected fluences at T=10 $K$. (c) The fluence dependence of the $\Delta I_{SHG}/ I_{SHG}$  at time delay $\tau$= 10 ps (the black circles) and the thermal effect $\Delta I_{SHG}/ I_{SHG}$ calculated at each fluence with respect to the pump-induced lattice temperature rise (the red squares).}\label{Fig:4}
\end{figure}

\textbf{The time-resolved second-harmonic generation (TR-SHG).} It is well known that the nonlinear susceptibility (especially second-order) is very sensitive to the symmetry change of a lattice \cite{SHG2005,PRLBi2Se3,BykovBi2Se3,GaAs2003}. As demonstrated in Fig. \ref{Fig:1} (d), the SH intensity in MoTe$_2$ is nearly zero in the 1T$^{'}$ phase while sharply enhanced in the T$_d$ phase where inversion symmetry is broken, hence more valuable insight about pump-induced lattice symmetry change can be gained from the TR-SHG spectroscopy. We performed a nir-infrared (800 nm) pump, SHG measurement on the sample. Earlier structural study based on crystallographic data analysis of single crystal samples has revealed that the polar direction related to the inversion symmetry breaking of MoTe$_2$  in the low temperature phase is along the c axis \cite{Sakaie1601378}. A normal incident beam from the regenerative amplifier with 1 K repetition rate is used as the pump excitation. The p-polarized SH photons with p-polarized incidence are recorded as the probe. Obviously, the change of the probe is proportional to the change of the polar degree (as demonstrated in Fig. \ref{Fig:1} (e) and (f)). Figure \ref{Fig:4} (a) shows the temperature-dependent relative change of SHG intensity at fixed pump fluence about 2 $mJ/cm^2$ as a function of time delay. We find that the SHG intensity has no response above T$_V$ while it drops to a lower value quickly (about 0.7 ps) upon pump excitation and lasts over 100 picoseconds below T$_V$. The SH intensity in the T$_d$ phase drops rapidly as temperature decreases and saturates at about 50 K. The 0.42 THz (period about 2 ps) phonon is also visible below 150 K in Fig. \ref{Fig:4} (a). Though the linear phonon spectroscopy also has detected the 0.42 THz shear mode, the mechanisms behind the dynamical oscillation through the linear and nonlinear probe are different. The former has been ascribed to the displacive excitation of coherent phonon mechanism which is linked to Raman tensor while the latter has often been taken as a hyper-Raman process\cite{GaAs2003}. The symmetry of the potential changes under intense fluence at low temperature was already demonstrated from the coherent phonon spectrum in Fig. \ref{Fig:3}. The reduction of the SH signal after intense pump implies that the polar degree of the crystal indeed decreases but not vanishes yet, namely the positions of the atoms have not yet shifted to the totally centrosymmetric phase. The fast decrease of signal intensity within about 0.7 ps after pumping suggests that the structural transition takes place in a subpicosecond time scale.

To explore the impact of pump fluence on the polar of the compound, we performed fluence-dependent time-resolved SHG measurements at T=10 K (Fig. \ref{Fig:4} (b)). We can see that the shape of the SHG transient does not change and the intensity drop is nonlinearly increased with the increase of the fluence (the empty black circles in Fig. \ref{Fig:4} (c)). The SH intensity drops to a maximum of nearly 80\% but does not completely disappear even when the fluence is close to the measured single-shot damage threshold. The 0.42 THz coherent phonon disappears at higher fluence. Here we would like to make a comparison between the thermal effect and the fluence dependent SH signal change. We calculate the pump-induced lattice temperature rise $T_{thermal}$ at each fluence based on the description presented in the above session incorporated with the temperature dependent $I(T)$ displayed in Fig. \ref{Fig:1} (d), then we can get the quantity $\frac{I(T_{thermal})-I(T_0)}{I(T_0)}$ as a function of $T_{thermal}$, being shown as red squares in Fig.\ref{Fig:4}(c)). We find that the fluence-dependent time-resolved SHG measurement roughly overlaps with the thermal effect plot, suggesting again an agreement with the lattice thermal effect.

\textbf{Discussions}
 The combined coherent phonon spectroscopy and the TR-SHG study reveal unambiguously a photo-induced switch between the topological nontrivial T$_d$ phase and a trivial 1T$^{'}$ phase in MoTe$_2$. Several features deserve to be further discussed and addressed. First, the time resolved SHG measurement demonstrates that the sharp drop of SHG signal is within a time scale of 1 ps after ultrafast excitations. This time scale is intrinsic since it is longer than the experimental time resolution. It is noted that the response time of SHG is comparable to the recovery time of electronic subsystem. Those observations indicate that the electronic and lattice degree of freedoms must have a very strong coupling, enabling a rapid transfer of energy from the electronic to lattice subsystem.
Second, both the pump-probe and the time-resolved SHG measurements reveal that the photo-induced state lasts for very long time without showing noticeable decay (see inset of Fig. \ref{Fig:4} (a)). This indicates that the intense photoexcitations actually drive the T$_d$  phase of the compound to a metastable phase, though this metastable phase is a known one in equilibrium, i.e. monoclinic 1T$^{'}$ phase. In fact, for many quantum materials, ultrashort pulses can drive them to certain novel quantum phases inaccessible at equilibrium.
Third, it is well known that the energy difference between T$_d$ and 1T$^{'}$ phase is small. Theoretical calculations \cite{Kim} suggested that the hole doping can destroy the orthorhombic phase and stabilize the monoclinic phase in MoTe$_2$ once the doped hole density exceeds $\sim10^{20} cm^{-3}$. Different from the hole doping, the ultrashort photo excitation presented in this work offers a new route to manipulate different orders in MoTe$_2$ .

The present study represents the first realization of switch between type-\uppercase\expandafter{\romannumeral2} Weyl semimetal T$_d$ phase and normal semimetal 1T$^{'}$ phase in bulk MoTe$_2$ among all possible tuning methods. It opens up new possibilities for ultrafast manipulation of the topological properties of solids, thus enabling potentially practical applications for topological switch device with ultrafast excitations. We remark that, although the photoinduced effect seems to be consistent with the lattice thermal effect, the direct structural change from a T$_d$ to 1T$^{'}$ phase induced by the intensive pump is another possibility for explaining the experimental observations. Our results indicate that investigating photoinduced structural change and its decay dynamics by direct time resolved structural probes, e.g. time-resolved X-ray diffraction, would be an interesting direction to pursue in future research.

\textbf{Methods}

\begin{small}
\textbf{Sample preparation and transport measurement}. The MoTe$_2$ single crystals were grown by vapor transport method using iodine as the transport agent, similar to the method reported in earlier work\cite{doi:10.1063/1.4947433}. Stoichiometric amounts of Mo and Te were first ground, pressed into pellets, and heated in an evacuated quartz tube at 810 $^{\circ}$C for 7 days to get polycrystalline MoTe$_2$ powder. Then, the powders and agent I$_2$ were mixed and sealed into another evacuated quartz tubes. The sealed quartz tubes were put in a two-zone tube furnace. The hot zone was kept at 1010 $^{\circ}$C and cold zone kept at 910 $^{\circ}$C and dwelled for 7 days. After that, the quartz tube was quenched into ice water to yield the high-temperature monoclinic phase. Plate-like shape single crystals with shinning surfaces were obtained. The resistivity was measured by a by a four-probe method and heat capacity measured by a heat-pulse relaxation method using Physical Properties Measurement System.

\textbf{Optical pump probe spectroscopy}. The transient reflectivity of the compound was measured with a standard pump-probe set up. A Ti:sapphire amplified laser (800-nm laser pulses with 35 fs duration at 1 kHz repetition rate) was utilized as the source of both pump and probe beams. The pump beam was modulated by a chopper at a frequency of 333 Hz and polarized in perpendicular to the probe beam. We kept the pump fluence at about 100 $\mu$J/cm$^2$, 10 times stronger than the probe pulses.

\textbf{Symmetry analyze}. In order to monitor the ultrafast dynamics of phase transition, the equilibrium technique can be extended to time domain via standard pump probe configuration, in which one stronger pulse trigger the sample and another probing pulse monitor the evolution of selected optical constant as a function of time delay. When the pulse duration of triggering femtosecond laser pulse is much shorter than the vibration period, then coherent motions of lattice, or the coherent phonon are induced, in which $<Q_{ks}>$ is nonzero and time dependent, oscillating with a frequency $\omega_{ks}$ with phase well defined with respect to an externally defined reference. In this coherent motion state, the dielectric susceptilbility can be expressed by displacement coordinate $Q(t)$ around the equilibrium as: $\chi=\chi_0+(\frac{\partial \chi}{\partial Q}*Q(t))$, where the second rank tensor $\frac{\partial \chi^{(1)}}{\partial Q}$ is linked to Raman tensor and third rank tensor $\frac{\partial \chi^{(2)}}{\partial Q}$ is related to hyper-Raman tensor. The coherent oscillations stem from the impulsively excited vibration of the A$_1$ phonon that leads to an isotropic modulation of refractive index vis the diagonal elements of the Raman tensor. To the first order in$Q(t)$, the reflectivity change $\Delta R(t)$ associated with a displacement $Q(t)$ can be expressed as: \par
$\Delta R(t) \propto \overrightarrow{E} ^{r}(\partial \chi / \partial Q) \overrightarrow{E} Q(t) $ \par
Here, $\overrightarrow{E} $ corresponds to the incident and $\overrightarrow{E} ^{r}$ to the reflected probe field. In our measurements, we only detect the intensity of reflected light instead of palarizition changes. Therefore, only the fully symmetry phonon, the diagonal elements of Raman tensor can be detected in our configuration.

\textbf{The time-resolved second-harmonic generation (TR-SHG) and SH polarization analysis}. The optical SHG was measured with 800 nm fundamental photons focused on the ab plane of the sample (0.1 to 0.8 mW on an ~80-mm spot) in reflection geometry at a 45$^{\circ}$ angle of incidence. The p (s) polarization of the incidence is in the ac-plane (along the b-axis) of the sample crystal. The incident polarization was controlled by a l/2 wave plate, and the polarization of the reflection SHG was analyzed by a Glan laser prism. The pump is in normal incidence and the scattering surface is in the ac-plane of the compound.

Without pump, the signal is absent at high temperature but increases suddenly below the phase transition as shown in Fig.\ref{Fig:1} (d). The polarization dependence of the SHG signal can be understood by considering the nonlinear properties of the point group $mm$2 as predicted for MoTe$_2$. For Fig.\ref{Fig:1} (f), the incident electric field $\hat{e}_{\omega}^{\phantom{12}2}=[E_{x}^{2},  E_{y}^{2}, E_{z}^{2}, 2E_{y}E_{z}, 2E_{z}E_{x},2E_{x}E_{y}]=[\frac{1}{2}cos^{2}\theta ,sin^{2}\theta,\frac{1}{2}cos^{2}\theta,-\sqrt{2}sin\theta cos\theta,- cos^{2}\theta,\sqrt{2}sin\theta cos\theta]$ ($\theta$ is the angle the incidence polarizer rotated from p polarization), generates a second harmonic signal along the $\hat{e}_{2\omega }=[0,1,0]$ for s polarization. The generated SHG intensity $I_{SHG}$ can be expressed as $I_{SHG}\propto\left | \hat{e}_{2\omega }\cdot \textbf{d}:\hat{e}_{\omega}^{\phantom{12}2}\right |^{2}$, where $\textbf{d}$ is the contracted notation of the second-order susceptibility tensor. $\textbf{d}$ is a $3\times 6$ second-rank tensor and the non-zero tensor elements for the $mm$2 point group is $d_{15},d_{24},d_{31},d_{32}$ and $d_{33}$. So we can easily lead to $I_{s-out}\propto d_{24}^{\phantom{12}2}sin^22\theta$. It is used to fit the polarization pattern in Fig.\ref{Fig:1} (f) very well. Similarly, when $\theta =0$, the incident electric field is $\hat{e}_{\omega}^{\phantom{12}2}=[E_{x}^{2},  E_{y}^{2}, E_{z}^{2}, 2E_{y}E_{z},  2E_{z}E_{x},2E_{x}E_{y}]=[\frac{1}{2},0,\frac{1}{2},0,-1,0]$. We fit the lobes in Fig.\ref{Fig:1} (e) by equation $I_{p-in} \propto (-d_{15}+d_{31}/2+d_{33}/2)^2 cos^2\varphi$, where SH analyzer rotated at an angle $\varphi$ from p polarization.

\end{small}

\bibliographystyle{nc}
  \bibliography{mote2}

\begin{small}
 {\textbf{Acknowledgement}}\par
We acknowledge useful help from Dr. H. P. Wang in the data analysis. This work was supported by the National Science Foundation of China (No. 11327806, 11404385 GZ1123, No. 11774008), and the National Key Research and Development Program of China (No.2016YFA0300902, 2017YFA0302904, Grant No. 2018YFA0305604 and No. 2017YFA0303302), the Key Research Program of the Chinese Academy of Sciences (Grant No. XDPB08-2).

{\textbf{Author contribution}}\par
 All authors contributed to the intellectual contents of this work. N. L. Wang and T. Dong designed the experiments. M. Y. Zhang and T. Dong developed the time resolved second harmonic generation spectrometer. Y. N. Li and J. Wang prepared the sample and the resistivity measurement. M. Y. Zhang,Y. N. Li and L. Y. Shi collected the data. M. Y. Zhang,Z. X. Wang and T. Dong contributed to the symmetry analysis. M. Y Zhang, T. Dong and N. L. Wang interpreted the data and wrote this paper. All authors participated in scientific discussions.

{\textbf{Competing financial interests}}\par
The authors declare no competing financial interests.

\end{small}

\end{document}